\documentclass[prl,twocolumn,superscriptaddress,floatfix,noshowpacs,notitlepage]{revtex4-1}%
\usepackage{graphicx,bm,times}
\usepackage{amsmath}
\usepackage{amsfonts}
\usepackage{amssymb}

\begin{document}

\title{Evidence for unidirectional nematic bond ordering in FeSe}

\author{M. D. Watson}
\email[corresponding author:]{matthew.watson@diamond.ac.uk}
\affiliation{Diamond Light Source, Harwell Campus, Didcot, OX11 0DE, UK}

\author{T. K. Kim}
\affiliation{Diamond Light Source, Harwell Campus, Didcot, OX11 0DE, UK}

\author{L. C. Rhodes}
\affiliation{Department of Physics, Royal Holloway, University of
	London, Egham, Surrey TW20 0EX, UK}

\author{M. Eschrig}
\affiliation{Department of Physics, Royal Holloway, University of
	London, Egham, Surrey TW20 0EX, UK}

\author{M. Hoesch}
\affiliation{Diamond Light Source, Harwell Campus, Didcot, OX11 0DE, UK}

\author{A. A. Haghighirad}
\affiliation{Clarendon Laboratory, Department of Physics,
	University of Oxford, Parks Road, Oxford OX1 3PU, UK}

\author{A. I. Coldea}
\affiliation{Clarendon Laboratory, Department of Physics,
	University of Oxford, Parks Road, Oxford OX1 3PU, UK}

\begin{abstract}

The lifting of $d_{xz}$-$d_{yz}$ orbital degeneracy is often considered a hallmark of the nematic phase of Fe-based superconductors, including FeSe, but its origin is not yet understood. Here we report a high resolution Angle-Resolved Photoemission Spectroscopy study of single crystals of FeSe, accounting for the photon-energy dependence and making a detailed analysis of the temperature dependence. We find that the hole pocket undergoes a fourfold-symmetry-breaking distortion in the nematic phase below 90~K, but in contrast the changes to the electron pockets do not require fourfold symmetry-breaking. Instead, there is an additional separation of the existing $d_{xy}$ and $d_{xz/yz}$ bands - which themselves are not split within resolution. These observations lead us to propose a new scenario of ``unidirectional nematic bond ordering" to describe the low-temperature electronic structure of FeSe, supported by a good agreement with 10-orbital tight binding model calculations.

\end{abstract}
\maketitle




The search for the understanding of unconventional superconductivity in the Fe-based systems has lead to a focus on the origin and nature of the ordered phases found in close proximity to the superconducting phase. Particular attention has been drawn to the ``nematic" phase \cite{Fernandes2014}, where the four-fold symmetry of the lattice is broken at a temperature $T_s$ which is higher than the striped antiferromagnetic ordering temperature $T_N$ in some materials \cite{Kim2011}. There has been long discussion about whether these two transitions are both magnetic in origin, or whether $T_s$ corresponds to a separate orbital instability \cite{Fernandes2014}. FeSe is an exceptional case, since it undergoes a structural transition at $T_s$~=~90~K without long range magnetic order at any temperature, enabling detailed study of the symmetry-broken state \cite{Bohmer2013,Bohmer2014correct,Baek2014,Rahn2015,Wang2016,Wang2015c_arxiv,Shamoto2015_arxiv,Watson2015b}, and has therefore attracted much theoretical  attention \cite{Chubukov2015,Glasbrenner2015,Yu2015a,Mukherjee2015,Yamakawa2015,Wang2015a,Lovesey2016}. 


Angle-Resolved Photoemission Spectroscopy (ARPES) measurements of FeSe \cite{Tan2013,Maletz2014,Shimojima2014,Nakayama2014,Watson2015a,Zhang2015,Watson2015c,Suzuki2015,Borisenko2015,Li2015} provide direct experimental access to the evolution of the electronic structure through $T_s$, which can be linked to theoretical models of the underlying symmetry-breaking order. Most previous ARPES studies have paid particular attention to the electron pockets around the M point, and inferred a $\sim$50~meV splitting of $d_{xz}$ and $d_{yz}$ bands, similar to previous claims in NaFeAs \cite{Yi2012} and Ba(Fe$_{1-x}$Co$_x$)$_2$As$_2$ \cite{Yi2011}. In this scenario, which could be interpreted as a ferro-orbital ordering \cite{Watson2015a}, the bands of primarily $d_{xz}$ and $d_{yz}$ character are degenerate exactly at M in the high temperature phase \cite{Fernandes2014b}, but split below $T_s$. Then, the dispersions observed by ARPES all arise from $d_{xz}$ or $d_{yz}$ bands, taking into account the presence of twin domains in the sample, with the assumption that outer electron band with $d_{xy}$ character does not contribute \cite{Watson2015a}. While alternatives to the ferro-orbital ordering scenario have been recently proposed \cite{Chubukov2015,Suzuki2015,Liang2015,Jiang2016,Onari2015_arxiv}, the splitting of $d_{xz}$ and $d_{yz}$ bands at the M point has been considered a hallmark of the nematic phase, until now \cite{Nakayama2014,Shimojima2014,Watson2015a,Watson2015c}.   

In this letter, we present a high-resolution ARPES study of the evolution of the electronic structure of bulk FeSe through the nematic transition at $T_s$. We use curvature analysis and a new procedure to fit the Energy Dispersion Curves (EDCs) at the M point to extract band positions. We also observe the $k_z$ dependence of the electron pockets, and present new Fermi surface maps covering the whole Brillouin zone above and below $T_s$. While the hole pockets of FeSe do undergo significant symmetry-breaking distortions below $T_s$, the changes to the electron pockets arise from an additional separation of existing $d_{xz/yz}$ and $d_{xy}$ bands, without any resolvable lifting of $d_{xz}$-$d_{yz}$ band degeneracy at the M point. These observations exclude the ferro-orbital scenario, and place new strong constraints on the nature of the orbital ordering. Finally we use a ten-orbital tight binding model to propose that the nematic state of FeSe is characterised by rotational-symmetry breaking within Fe-Fe hoppings, which may be described as a unidirectional nematic bond ordering. 

High-quality single crystal samples of FeSe were grown by the chemical vapor transport method \cite{Chareev2013,Bohmer2013,Watson2015a}. ARPES measurements were performed at the I05 beamline at the Diamond Light Source, UK. 

\begin{figure*}[t!]
	\centering
	\includegraphics[width=0.95\linewidth]{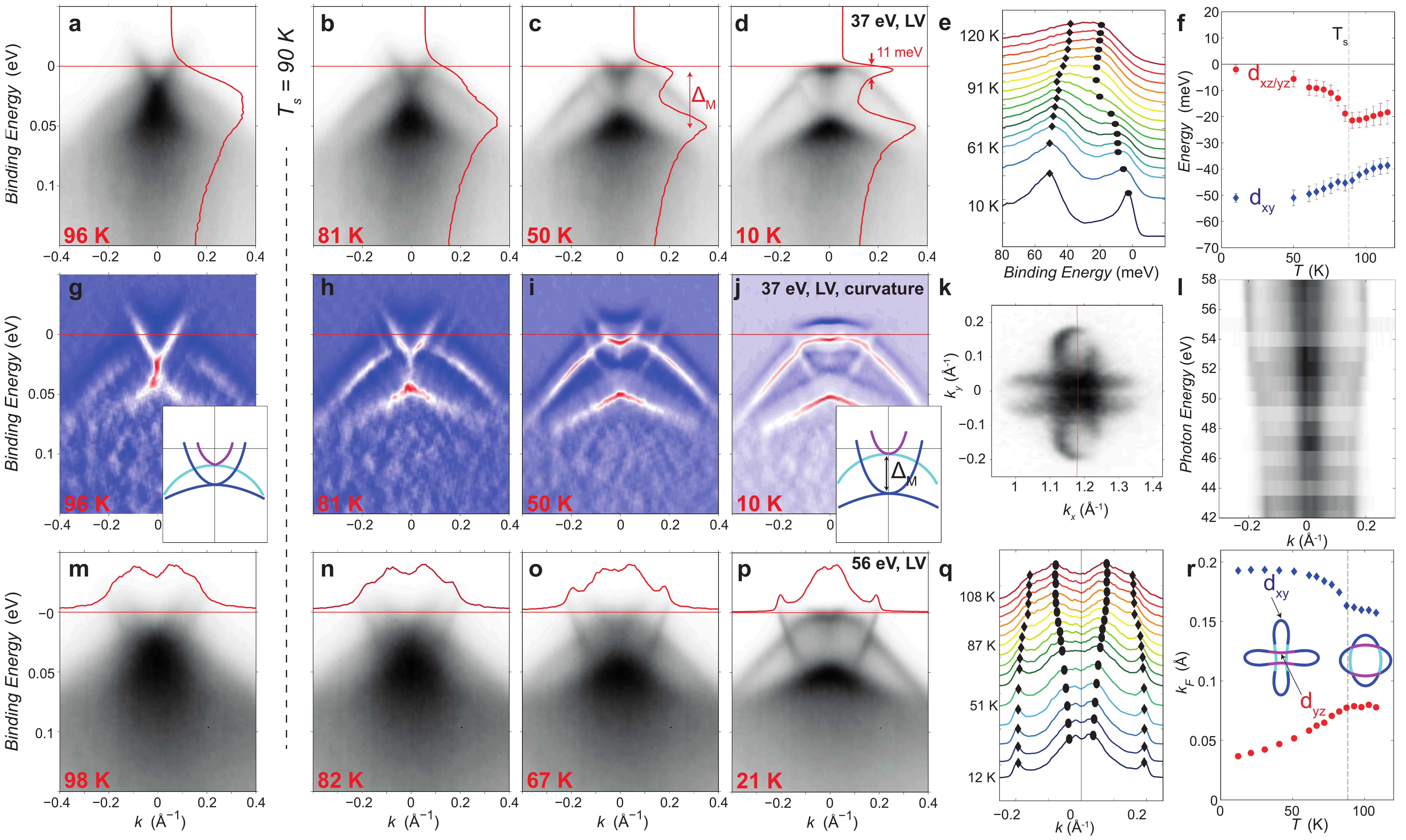}
	\caption[~]{a-d) ARPES intensity plots of a high-symmetry cut through the M point, using 37 eV photon energy in vertical polarisation (LV- perpendicular to scattering plane). Solid lines plot the EDC at M, which are also stacked according to temperature in e). f) Temperature-dependence of peak positions extracted from the EDC analysis described in the text. g-j) Curvature plots, showing traces of all expected bands (as shown in cartoon insets) above and below $T_s$. k) Low temperature electron Fermi surface at 56 eV. l) Photon energy dependence of the MDC through the M point at 10~K, which corresponds to the $k_z$ dispersion of the electron pocket. The A point where pockets are largest corresponds to 56 eV. m-p) High symmetry cuts (along red line in k)) through the A point at 56 eV, showing also the MDC at the Fermi level. q) MDCs plotted as a function of temperature and r) temperature-dependence of $k_F$ vectors extracted from fits to MDCs.}
	\label{fig:fig1}
\end{figure*}

In Fig.~\ref{fig:fig1}a-d) we present temperature-dependent ARPES data for a high symmetry cut centred on the M point, also showing the EDCs at M. At low temperatures (Fig.~\ref{fig:fig1}d)), these EDCs display two prominent peaks with a separation $\Delta_M$~=~50~meV, previously attributed to orbital splitting in the nematic phase, i.e. $\Delta^{FO}_M=E_{yz}-E_{xz}$. However, curvature analysis \cite{Zhang2011_curvature} enhances weak features in the data and provides a different perspective, as shown in Fig.~\ref{fig:fig1}g-j). Above $T_s$, in Fig.~\ref{fig:fig1}g) both the expected $d_{xz}$ and $d_{yz}$ dispersions and also sections with $d_{xy}$ character are observed. By comparing Fig.~\ref{fig:fig1}g-j), we observe the similarity in the dispersions above and below $T_s$, and no extra features arise. Therefore here we propose a new scenario, where the low temperature dispersions simply consists of the $d_{xz/yz}$ and $d_{xy}$ bands as expected above $T_s$, but with an increased separation between them. We assign the peaks in the EDCs to the two bands, i.e. $\Delta_M=E_{xz/yz}-E_{xy}$, equal to 50 meV at low temperatures but also finite above $T_s$. Within experimental resolution (11 meV FWHM for the relevant peak in Fig.~\ref{fig:fig1}d)) we do not detect any subsequent splitting between $d_{xz}$ and $d_{yz}$ bands, although since this is allowed by symmetry in the orthorhombic phase a small splitting may occur as a secondary effect.

 \begin{figure*}[t!]
 	\centering
 	\includegraphics[width=0.95\linewidth]{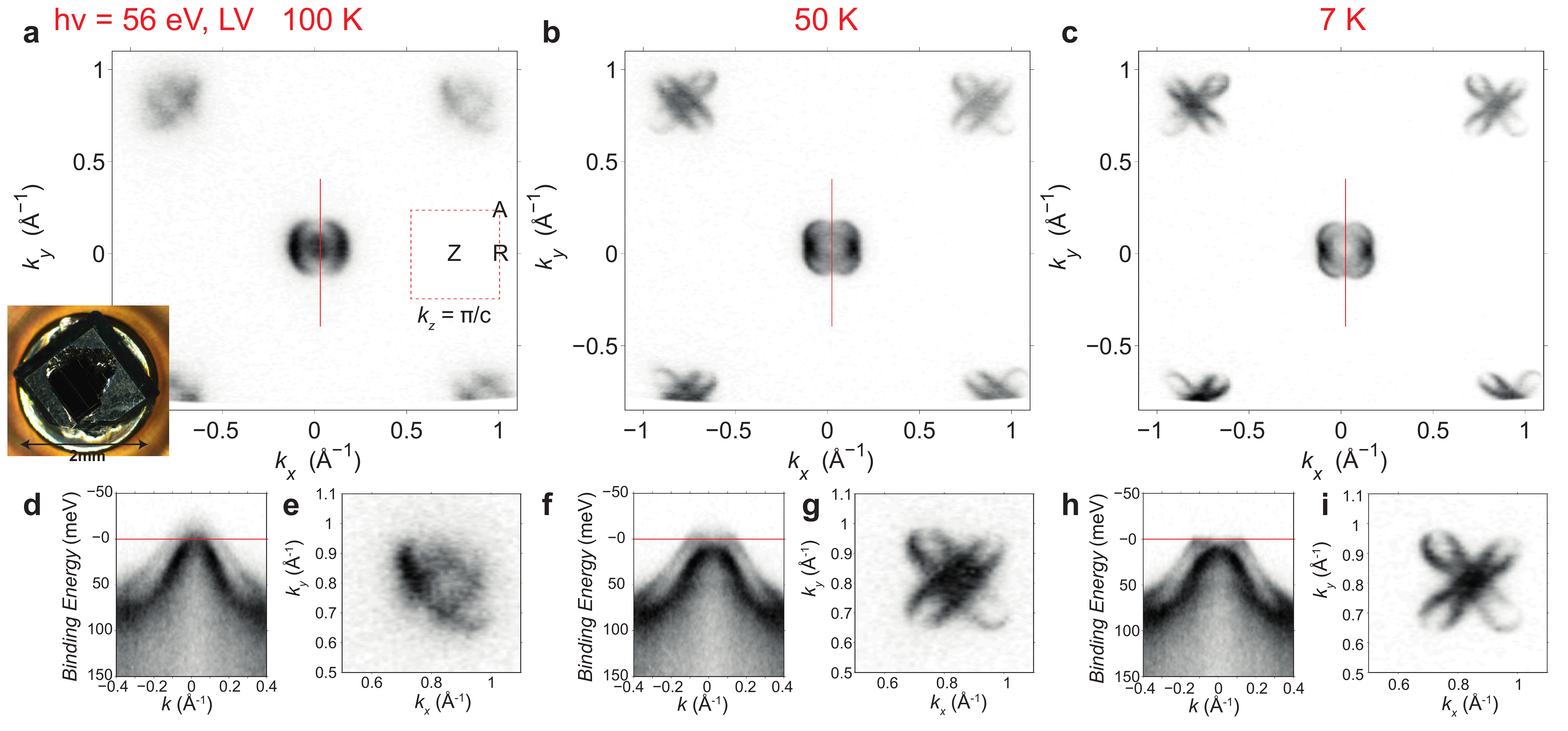}
 	\caption[~]{a-c) Temperature-dependent Fermi surface maps of FeSe near the top of the Brillouin zone, taken using 56 eV photon energy in LV polarisation and integrating spectral weight within 2 meV of the Fermi level. Inset to a) is the sample used. d,f,h) Cuts through the Z point (RZR direction - solid red line in maps a-c), revealing the extra splitting which pushes the inner hole band below the Fermi level. e,g,i) Expanded Fermi surface map of the electron pocket: both the inner and outer electron bands are clearly detected.}
 	\label{fig:fig2}
 \end{figure*}

This new scenario can be tested by the extraction of band positions as a function of temperature from the EDCs at the M point. However while the separation $\Delta_M$ of the two features in the EDC is unambiguous at low temperatures, it becomes more difficult to define at higher temperatures, as features become broader. Here we take a new approach: we fit the EDC with two asymmetric pseudo-Voigt functions and the Fermi function at 61~K where the peaks are still distinct. Then, at higher temperatures we fit using the same peak profiles, only allowing the peak positions to vary (see Supplemental Materials, SM). We find that the fitted peak positions are separated by 20 meV even above $T_s$, but increase substantially when the system enters the nematic phase (Fig.~\ref{fig:fig1}e,f). The sharp upward shift of the $d_{xz/yz}$ bands at $T_s$ is the strongest feature, although the $d_{xy}$ band position also adjusts downwards below $T_s$, leading to a total separation of 50 meV at 10~K. 

In Fig.~\ref{fig:fig1}m-p) we present new measurements of the electron pockets using a photon energy of 56 eV, where the outer $d_{xy}$ electron dispersion is already clearly visible in the data above $T_s$ even without curvature analysis. The 56 eV photon energy is chosen to correspond to the A point at the top of the Brillouin zone where the warped quasi-2D electron pocket is largest, giving the best momentum-resolution of features, as can be seen in Fig.~\ref{fig:fig1}l). Further details of the photon-energy dependence are presented in SM, where it is also shown that $\Delta_M$ is independent of $k_z$. In Fig.~\ref{fig:fig1}q,r) we extract the temperature-dependence of both $k_F$ vectors observed, from peak fitting of the Momentum Distribution Curve (MDC) at the Fermi level (SM). The $k_F$ values are well-defined at all temperatures and also demonstrate that the $d_{yz}$ and $d_{xy}$ dispersions are separate above $T_s$ and undergo additional separation below $T_s$. 

In Fig.~\ref{fig:fig2} we present further support for the new interpretation including the observation of the $d_{xy}$ electron band. These ARPES data are obtained with the scattering plane at 45$^{\circ}$ to the Fe square lattice to mitigate matrix elements effects \cite{Brouet2012}, again using 56 eV photon energy. At 100~K, in the tetragonal phase above $T_s$, the Fermi surface maps (Fig.~\ref{fig:fig2}a,e) reveal essentially the whole structure of the expected electron pockets at the A point, including the outer pocket with $d_{xy}$ character. Since all expected bands contribute at high temperature, we conclude that the whole structure of the low temperature Fermi surface is also observed. The comparison between Fig.~\ref{fig:fig2}e) and i) demonstrates that the low-temperature Fermi surface is elongated compared to the high temperature case, but retains the same basic structure and symmetry. Thus the electron pockets retain fourfold symmetry and the bands from both structural domains will overlap in experiments.

The evolution of the electron pockets through $T_s$ contrasts strongly with the behaviour of the hole pockets. At the Z point (Fig.~\ref{fig:fig2}d), above $T_s$ the inner hole pocket just crosses the Fermi level, making a small 3D pocket \cite{Watson2015a,Watson2015c}, and there is a $\sim$20~meV splitting due to spin-orbit coupling at the Z point between the $d_{xz/yz}$ bands. However below $T_s$ there is an extra splitting of $\sim$15 meV associated with a $d_{xz/yz}$ orbital polarisation, in addition to the spin-orbit splitting. This pushes the inner band completely below the Fermi level (Fig.~\ref{fig:fig2}h). Therefore there is a single hole band at low temperatures which displays an elliptical distortion, however due to the presence of twin domains below $T_s$, two crossed ellipses are superposed in the ARPES data in Fig.~\ref{fig:fig2}b,c).

\begin{figure}
\centering
\includegraphics[width=0.85\linewidth]{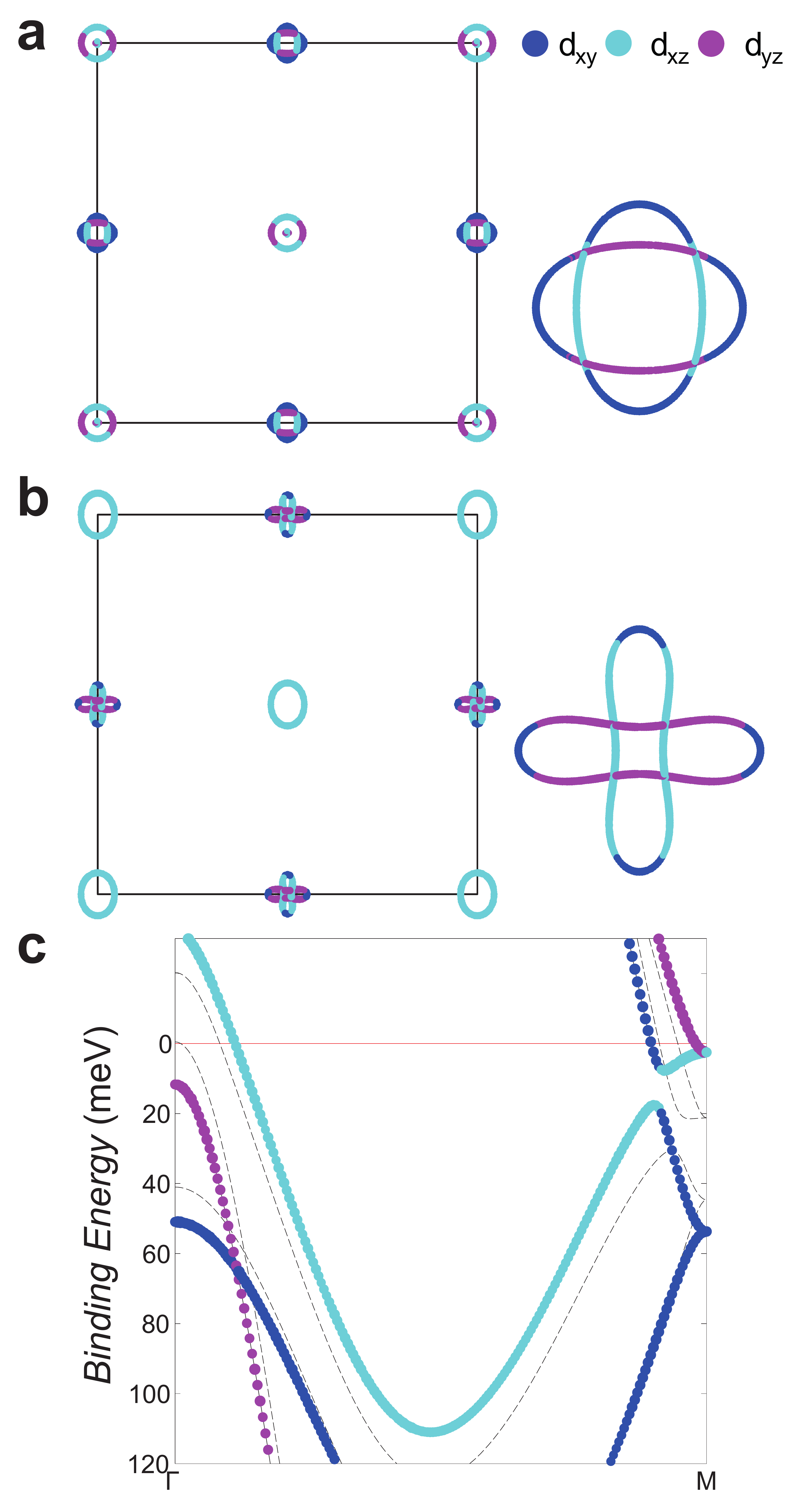}
\caption{a) Fermi surface calculated using a ten orbital tight binding model for FeSe in the tetragonal phase, showing the folded band structure, and b) with the unidirectional nematic bond order parameter, which distorts the hole pocket but preserves fourfold symmetry of the electron pocket. c) Band structure along $\Gamma$-M ((0,0)-(0,$\pi$) direction) with the dashed lines corresponding to the tetragonal phase. Note the extra splitting at the $\Gamma$ point and the symmetric shift of $d_{xz/yz}$ bands at the M point.}
\label{fig:fig3}
\end{figure}

The breaking of fourfold rotational symmetry of the hole pockets below $T_s$ is therefore well-established. However, we have found symmetry-preserving changes to the electronic structure at the M point, which on first sight are difficult to reconcile with the tetragonal-to-orthorhombic phase transition at $T_s$. Another constraint is that back-folded bands are not observed, suggesting that translational symmetry is preserved at $T_s$. The challenge is to identify an orbital order parameter which globally breaks fourfold rotational symmetry but is consistent with the strong constraints provided by these observations. Ferro-orbital ordering could account for the symmetry-breaking at the $\Gamma$ point, but requires a splitting of $d_{xz/yz}$ orbitals at the M point, which we have argued is not the case (see also SM). Moreover, ferro-orbital ordering is not consistent with the direction of distortion of the hole band, as revealed by ARPES measurements on detwinned crystals \cite{Suzuki2015}. A N\'{e}el-type antiferro-orbital ordering would preserve the $d_{xz/yz}$ degeneracy at M, but it cannot explain the extra splitting observed between $d_{xz/yz}$ bands at the $\Gamma$/Z point \cite{Liang2015}. The recently proposed $d$-wave bond nematic order predicts a splitting at the M point \cite{Liang2015,Jiang2016}, as does the microscopic model of Ref.~\cite{Onari2015_arxiv}. However, we suggest that a ``unidirectional nematic bond ordering" is compatible with all the observations. 

In  Fig.~\ref{fig:fig3} we present tight-binding model electronic structures with and without the proposed unidirectional nematic bond order. We use a 2D ten-orbital tight binding model including spin-orbit coupling \cite{Mukherjee2015,Eschrig2009} (SM). Within this model, we add an order parameter to the inter-site hopping block of the Hamiltonian as $h=\Delta_{S}(n'_{yz}-n'_{xz})\cos(k_x)$, where $n'_{xz}$ indicates the inter-site hopping operator distinct from on-site occupations $n_{xz}$. This order parameter therefore describes a symmetry-breaking within the inter-site $d_{xz/yz}$ hopping terms on the Fe-Fe bonds in the $x$ direction. This order parameter has the desired properties of giving an extra splitting in addition to the spin-orbit coupling splitting of $d_{xz/yz}$ bands at the $\Gamma$ point, but a symmetric shift up of the $d_{xz/yz}$ bands at the M point without losing the degeneracy. It still globally breaks fourfold rotational symmetry yet it can also be simply shown that it does not break translational symmetry (SM). In Fig.~\ref{fig:fig3}b,c) we present the results of a calculation with $\Delta_{S}$~=~20~meV; we also include an adjustment of -10~meV to the $d_{xy}$ orbital which is motivated by the experimental results in Fig.~\ref{fig:fig1} and may be required to maintain the charge balance of the system. Within this fairly simple model we obtain Fermi surfaces and dispersions which reproduce all qualitative features of the low-lying electronic structure. Therefore we suggest that the changes to the electronic structure of FeSe in the orthorhombic phase may be primarily described by a unidirectional nematic bond ordering.

Since the Fermi surface changes at the M point are not of a symmetry-breaking nature, and $\Delta_M$ is finite even above $T_s$, there is an important question about how tightly the evolution of the electron pockets is linked to $T_s$. However we have shown that there is a sharp increase in $\Delta_M$ which onsets exactly at $T_s$ (Fig.~\ref{fig:fig1}j). Additionally, the deviation in $k_F$ values of both the inner and outer electron pocket branches also follows a sharp order-parameter-like behaviour which onsets at $T_s$, and this deviation of the inner pocket was shown to behave as an order parameter of the structural transition across the FeSe$_{1-x}$S$_x$ series \cite{Watson2015c}. This indicates that the changes at the M point are still fundamentally linked to the orthorhombic lattice distortion. This may be understood since in the unfolded 1-Fe Brillouin zone, the electron pockets with $d_{xz}$ and $d_{yz}$ character are located in different parts of $k$-space, and therefore the $\cos{(k_x)}$ term ensures that they shift symmetrically, such that degeneracy is not lost in the folded 2-Fe zone (SM). Therefore the apparently non-symmetry breaking band movements at M are linked to an ordering which globally \textit{does} break tetragonal symmetry concomitant with the tetragonal-orthorhombic lattice distortion at $T_s$. 

Within the longstanding debate about the roles of orbital and spin degrees of freedom in Fe-based superconductors \cite{Fernandes2014}, FeSe is often considered as an example where orbital interactions may be dominant \cite{Bohmer2013,Bohmer2014correct,Baek2014,Shamoto2015_arxiv,Watson2015a}. However we have excluded any significant ferro-orbital ordering, and furthermore we have found that the primary order parameter is bond-centered and that translational symmetry is not broken, in contrast to some proposals \cite{Chubukov2015,Kontani2011}. Intriguingly, a bond-centered ordering is also observed in cuprates \cite{Comin2015}, although in that case it has an incommensurate modulation. On the other hand, there are mixed reports on how relevant magnetic interactions are to the structural transition in FeSe \cite{Bohmer2013,Bohmer2014correct,Baek2014,Rahn2015,Wang2016,Wang2015c_arxiv,Shamoto2015_arxiv,Glasbrenner2015}. The magnitude of energy shifts (e.g. 20 meV for the $d_{xz/yz}$ bands at M) are similar to the spin-orbit coupling value of 20 meV in FeSe \cite{Watson2015a}, suggesting that a spin-driven scenario is not ruled out. Thus our data excludes several proposed orbital ordering scenarios, and instead points to the existence of a unidirectional nematic bond ordered phase in FeSe, breaking rotational but not translational symmetry, which is distinct from the known striped antiferromagnetic phase in other Fe-based superconductors. 

In conclusion, we have presented a high-resolution ARPES study of FeSe, and argued that the symmetry-breaking distortion of the hole pockets at the $\Gamma$ point but symmetry-preserving changes of the electron pockets at the M point below $T_s$ point can be explained by a unidirectional nematic bond ordering. These measurements provide a fresh perspective on the nature of the nematic phase of Fe-based superconductors. 

\section{acknowledgments}

\begin{acknowledgments}
We thank T. Scaffidi, R. Valenti, M. Rahn, Z. Wang, R. Fernandes and F. Kruger for useful discussions. We thank N.~B.~M.~Schr\"{o}ter, L.~B.~Duffy, S. F. Blake and A. Narayanan for technical assistance.  We thank Diamond Light Source for access to Beamline I05 (Proposals Nos. SI10203,SI11792,SI12799) that contributed to the results presented here. Part of the work was supported by the EPSRC (EP/L001772/1,EP/I004475/1,EP/I017836/1). A.A.H. acknowledges the financial support of the Oxford Quantum Materials Platform Grant (EP/M020517/1). A.I.C. acknowledges an EPSRC Career Acceleration Fellowship (Grant No. EP/I004475/1).

\end{acknowledgments}

\clearpage
\section{Supplemental Material}

\subsection{Fitting of EDCs at the M point}

\begin{figure}[h!]
	\centering
	\includegraphics[width=\linewidth]{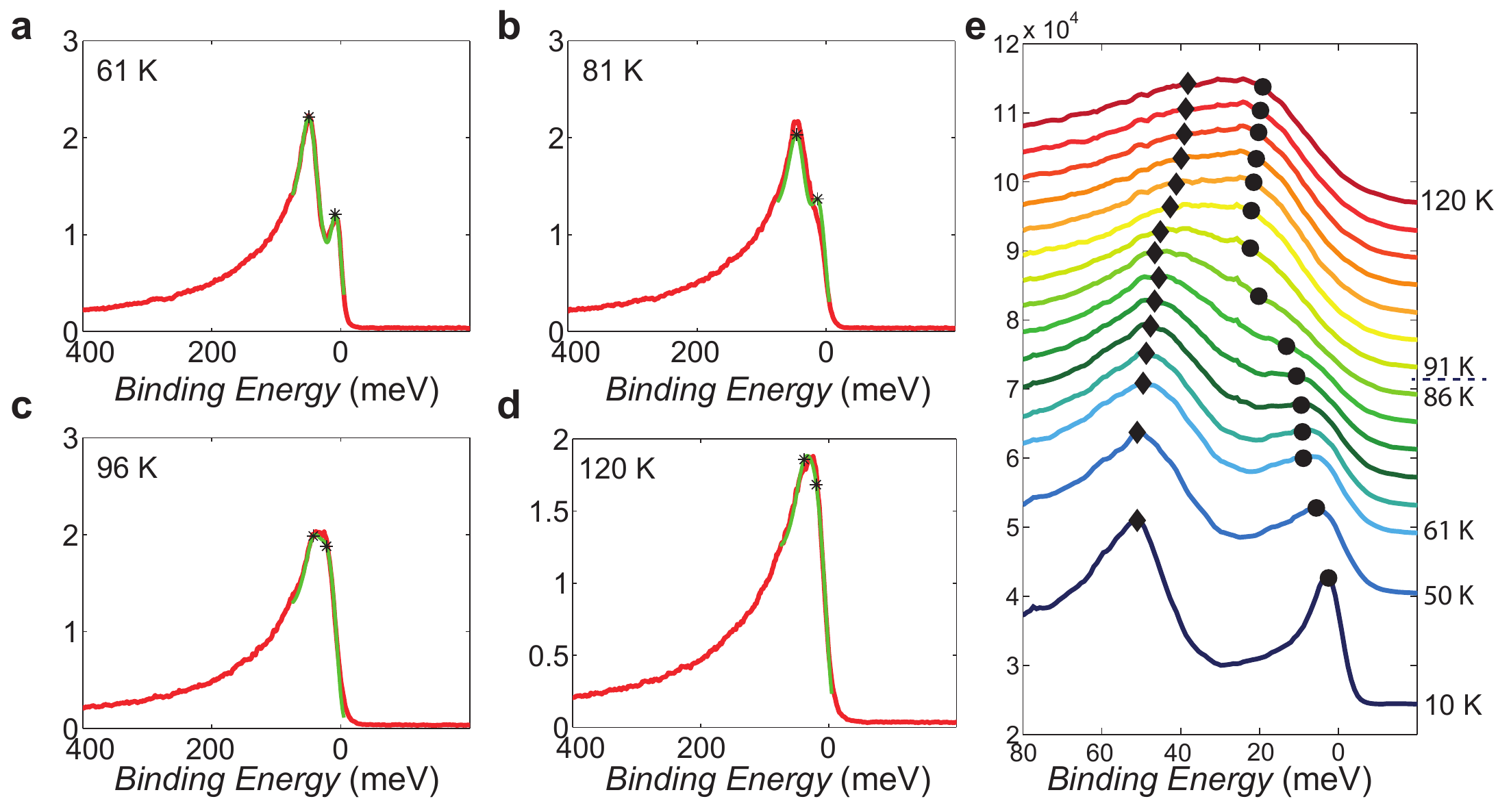}
	\caption{a-d) Representative fits of EDCs following the procedure described in the main text. Black star symbols indicate the peak positions of the fit. e) Extracted peak positions plotted on top of raw EDC data, as a function of temperature.}
	\label{fig:sm1}
\end{figure}

The extraction of band positions at the M point has been an important part of previous studies of FeSe, where it has been claimed that a split peak structure arises from a single peak (i.e. the degenerate $d_{xz/yz}$ bands at M) either at $T_s$ \cite{Shimojima2014} or above $T_s$ \cite{Nakayama2014,Zhang2015}. In the main text we argued that this is not the correct scenario, and that the EDC above $T_s$ must arise from two separate features, even if they are not well-distinguished, and there is a well-defined transition at $T_s$. Other studies have typically used a second derivative analysis to determine the band positions, but this can give conflicting results \cite{Zhang2015,Nakayama2014}. As described in the main text, for this study we decided to take a novel approach to the problem by directly fitting the EDC to extract band positions.  

At 61 K (Fig.~\ref{fig:sm1}a), the peaks are clearly distinct, and here we perform a fit to the data with two asymmetric pseudo-voigt functions with additional background, multiplied by the Fermi function - that is, with 14 free parameters such that the peak profiles are very well fitted. However, once this fit is converged, at higher temperatures we use exactly the same parameters (amplitude, peak width, asymmetry parameters and background), except for an overall peak width term which is fixed to increase linearly with temperature to account for the experimental increase in peak width with temperature. We therefore extract the band positions above 61~K from a final fit which has only the peak positions free - i.e. a fit with only two free parameters. Representative fits are displayed in Fig.~\ref{fig:sm1}. We decided that the slight loss in fit quality is compensated by the greater confidence in the peak positions when these are the only free parameters. Indeed, this procedure gives a sensible and systematic temperature-dependence of band positions as shown in Fig.~1j of the main text, and is complementary to the result of the MDC fitting analysis. Note that this fitting method in principle could converge with both peaks at the same position and summed intensity above $T_s$ which would be compatible with ferroorbital ordering and is therefore unbiased between the two interpretations, but in fact the method converges to be consistent with the revised interpretation with a 20 meV splitting at the M point at higher temperatures between $d_{xz/yz}$ and $d_{xy}$ bands. 

\subsection{Fitting of MDCs at M}

\begin{figure}[h!]
	\centering
	\includegraphics[width=\linewidth]{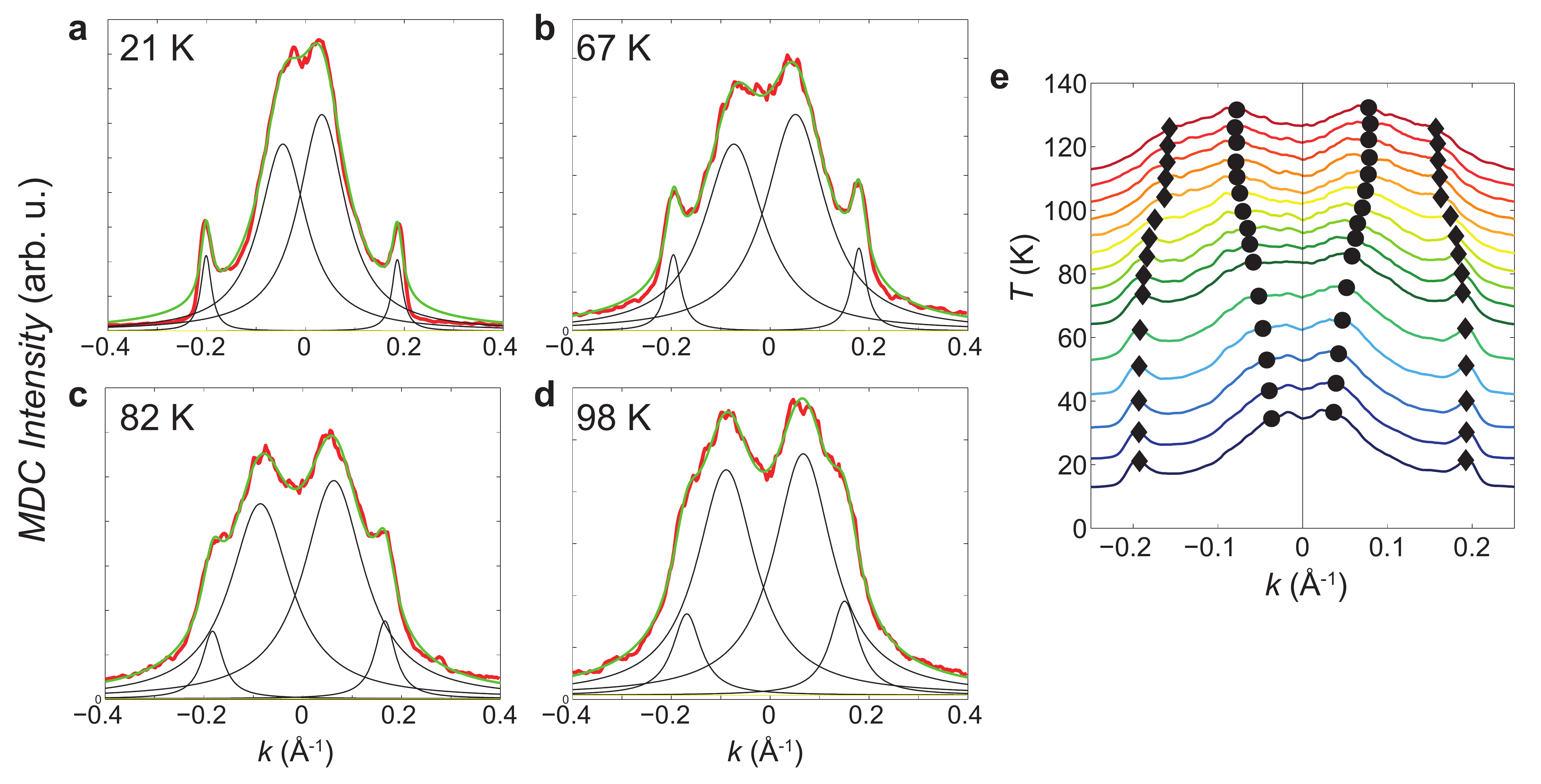}
	\caption{a-d) Representative fits of MDCs for the high-symmetry cut through the M point. Black lines represent the contributions of individual lorentzian peak profiles. e) Extracted $k_F$ values plotted on raw MDC data stacked according to temperature.}
	\label{fig:sm2}
\end{figure}

In Fig.~\ref{fig:sm2} we present MDC fit profiles of the high-symmetry cut through the M point at 56 eV, at representative temperatures, in order to show how the temperature-dependence of $k_F$ was determined in Fig.~1p) of the main text. For the fit procedure, we use two pairs of Lorentzian peaks with equal widths. We presented a  similar analysis using 37~eV data in Refs.~\cite{Watson2015a,Watson2015c}. However at that photon energy the outer branch has much weaker intensity (Fig.~1 of the main text) and cannot be tracked up to $T_s$.

\begin{figure*}
	\centering
	\includegraphics[width=0.7\linewidth]{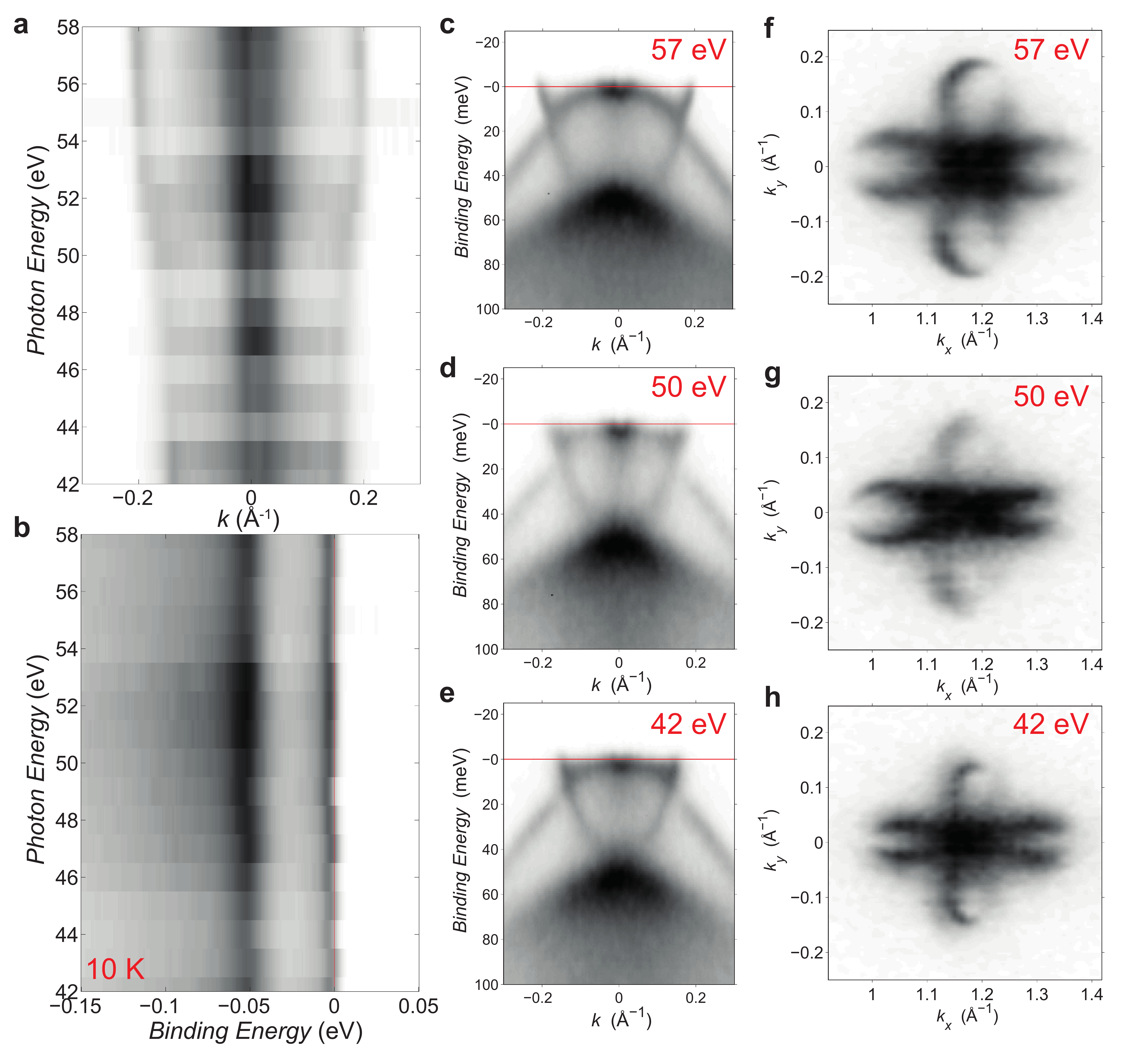}
	\caption{a) Photon-Energy dependence of MDCs (reproduced from Fig 2 of main text). b) Photon-energy dependence of the EDC through the M point, showing almost no deviation of the $\sim$50 meV splitting of bands at the M point as a function of $k_z$. c-e) Selected cuts used to construct a,b). The most significant difference as a function of $k_F$ is the dispersion of the outer electron band. f-h) Fermi surface mini-maps around the M point at selected photon energies. All measurements performed in Linear Vertical polarisation (normal to scattering plane).}
	\label{fig:sm3}
\end{figure*}

\subsection{Photon Energy dependence at M}
Here we present further details of the photon energy-dependence of the electron bands at M point, at low temperatures in a twinned sample. The photon energy dependence of the hole pockets of FeSe has been previously reported \cite{Watson2015a}, revealing a significant $k_z$ dependence of the outer hole pocket. At the M point, the Fermi surface shows a smaller warping effect, as the $k_F$ of the outer pocket shrinks from 0.193 \AA$^{-1}$ (56 eV - A point) to 0.142 \AA$^{-1}$ (42 eV - M point). This supports the analysis of quantum oscillation frequencies \cite{Watson2015a,Watson2015b} where it was determined that the electron pocket was less dispersive in $k_z$ compared to the hole pocket. It is interesting to note that the energy splitting at the M point appears to be $k_z$-independent, as shown in Fig.~\ref{fig:sm3}b). This indicates that the orbital ordering has no significant $k_z$ dependence. Thus the band positions at the M point do not change significantly with $k_z$, but the effective mass of the dispersion of the outer electron band from the M point does rise with $k_z$, giving rise to the warping of the Fermi surface. This can be directly observed by comparing the cuts at different photon energies shown in Fig.~\ref{fig:sm3}c-e), where the most substantial difference between them is the $d_{xy}$ band dispersion. Note that one previous measurement of the $k_z$ dependence at the M point did not find any significant $k_z$ dependence at the M-A point \cite{Li2015}, which we find surprising. Taking into account the $k_z$ dispersions of both the hole and electron pockets, we determine that the inner potential ($V_0$) appropriate for FeSe is approximately 12.2~eV.

\subsection{Tight binding model}
\begin{figure*}
	\centering
	\includegraphics[width=0.6\linewidth]{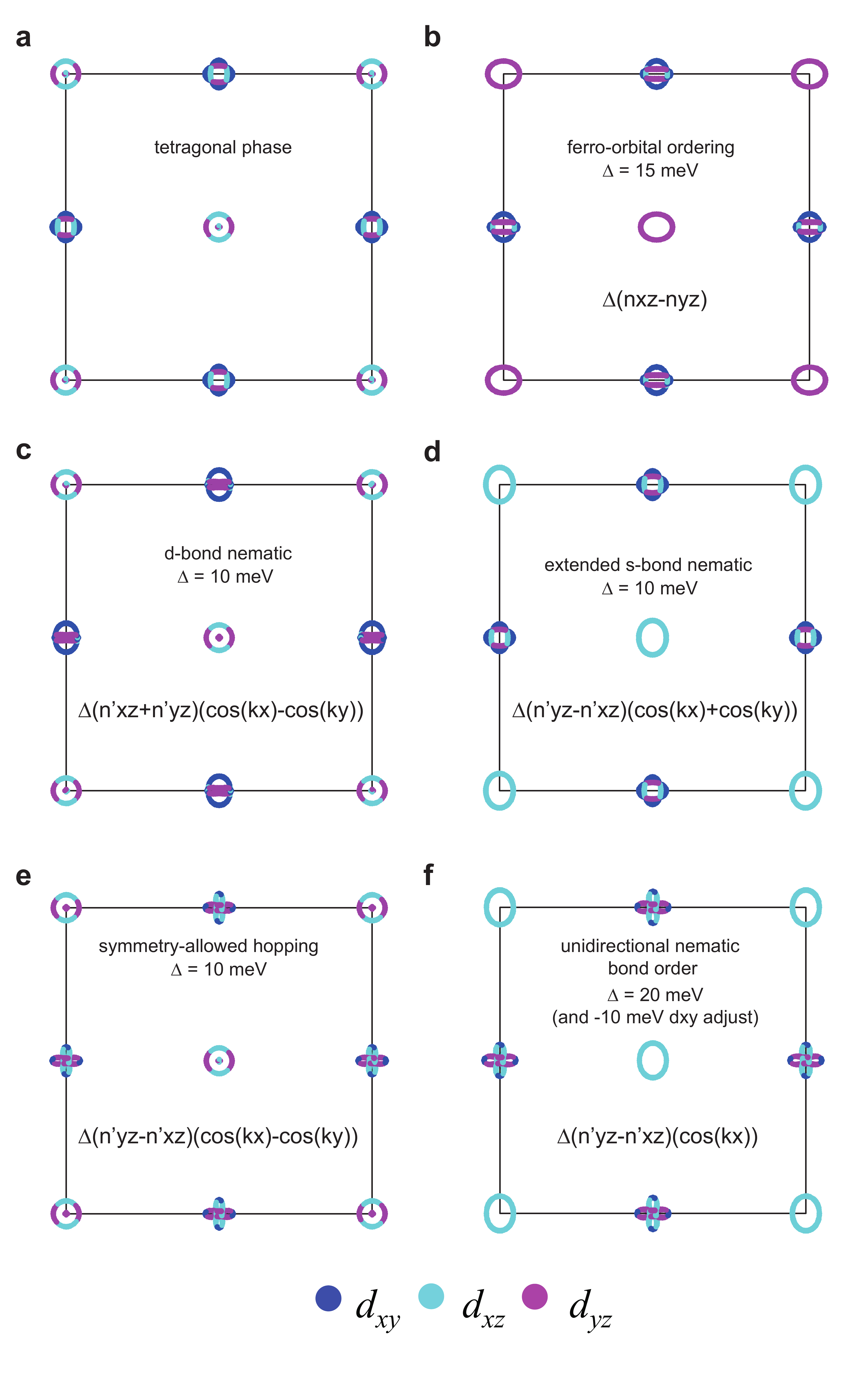}
	\caption{Fermi surface plots with different orbital ordering schemes from a 10-orbital tight binding model. Only the ``unidirectional nematic bond order" in f) can reproduce the whole experimental low-temperature Fermi surface.  }
	\label{fig:sm4}
\end{figure*}

\textit{The model: }In the main text, we presented Fermi surfaces and band dispersions for a 10-orbital tight binding model, in the form originally proposed in Ref.~\cite{Eschrig2009}. The parameters of Ref.~\cite{Mukherjee2015} were used as a starting point, with some adjustments made in order to better match the ARPES dispersions at high temperature. In practice we adapted the parameters until the low-lying bands at $\Gamma$ and M are described well; however some details such as the dispersion of the $d_{xy}$ hole band are less well captured. 

\textit{Motivation of order parameter: } We now discuss how the inclusion of a unidirectional nematic bond order term in our tight binding model is required in order to understand the ARPES data. Motivated by the experimental data, we added the following term to the 10-orbital model Hamiltonian in the inter-site blocks:

\begin{equation}
	h=\Delta_{S}(n'_{yz}-n'_{xz})\cos(k_x) .
	\label{eq:sm1}
\end{equation}

Here the $x$ direction corresponds to a Fe-Fe direction and the longer $a$ axis of the orthorhombic unit cell. We make the distinction here between $n_{xz}$ the on-site orbital number operator, and $n'_{xz}$ which is the operator for inter-site hopping between $d_{xz}$ orbitals. We describe it as ``unidirectional nematic bond order" since there is a $\cos(k_x)$ but not $\cos(k_y)$ term.

In the  unfolded 1-Fe unit cell (i.e. 5-orbital model), the two crossed ellipses that occur at the M point in the crystallographic 2-Fe unit cell are unfolded to different positions in momentum space; the pockets containing $d_{xz}$ and $d_{yz}$ character exist at $(0,\pm{}\pi)$ and $(\pm{}\pi,0)$ respectively. Experimentally, we have determined that these two ellipses must distort symmetrically. This could not occur for an order parameter defined homogeneously across the Brillouin zone, e.g. a ferro-orbital ordering, since the two pockets would distort in opposite directions due to their different orbital characters. Therefore we considered momentum-dependent order parameters, and in particular a $\cos(k_x)$ term ensures that the orbital splitting has different signs at $(\pi,0)$ and $(0,\pi)$. Due to the distinction in the orbital character at each electron pocket, this makes band shifts at each point occur symmetrically, and when folding back to the M point of the 2-Fe Brillouin zone, the Fermi surface has become more elongated as in experiment but the degeneracy of $d_{xz}-d_{yz}$ bands is not broken. A $\cos(k_y)$ momentum-dependence could also give symmetric shifts at the M point, however this would lead to the opposite distortions of the hole pocket compared to measurements on detwinned single crystals \cite{Suzuki2015}. Therefore we cannot rule out the existence of an additional bond nematic ordering with $\cos(k_y)$ modulation and significantly smaller magnitude than our primary $(n'_{yz}-n'_{xz})\cos{(k_x)}$ term. However experimentally the magnitude of the band shifts at M and the extra splitting at $\Gamma$ are similar and therefore a single unidirectional order parameter is sufficient.

\textit{Preservation of Translational Symmetry: }A simple argument can be made to show that this order parameter does not break translational symmetry. We can rewrite Eq.~\ref{eq:sm1} as:
\begin{equation}
	\begin{split}
		h=& \frac{\Delta_{S}}{2}(n'_{yz}-n'_{xz})(\cos(k_x)+\cos(k_y)) + \\  & \frac{\Delta_{S}}{2}(n'_{yz}-n'_{xz})(\cos(k_x)-\cos(k_y))
	\end{split}
\end{equation}

The first term here corresponds to ``extended $s$-wave bond nematic order" \cite{Jiang2016} and the second is an allowed hopping term which doesn't break any symmetries (see also Fig.~\ref{fig:sm4}). Therefore the symmetries which are broken at $T_s$ are those of the extended $s$-wave bond nematic order, i.e. breaking rotational but not translational symmetry (thus it is truly a ``nematic" order parameter). However although this decomposition is mathematically valid and noteworthy, it seems to us that our description of a unidirectional nematic bond order is more intuitively related to the underlying physics than using the description given by this linear combination. In particular, since the extended $s$-wave bond nematic order vanishes at the M point, this description would rely on symmetry-allowed hoppings to give the order-parameter-like behaviour at the M point \cite{Watson2015c}, which seems unphysical.

\textit{Comparison with other orbital orderings: }In Fig.~\ref{fig:sm4} we present a selection of Fermi surfaces in the presence of alternative orbital ordering schemes, sugggested by various authors for describing the nematic state. The simplest case is on-site ferro-orbital ordering \cite{Mukherjee2015} in b) which is not viable since it breaks $d_{xz}-d_{yz}$ symmetry at the M point. The $d$-wave bond nematic order \cite{Jiang2016,Liang2015} also fails due to the splitting at M and absence of splitting at $\Gamma$. The extended $s$-wave bond nematic order was introduced in \cite{Jiang2016} as a possible nematic order, but has no signature at the M point. In e) we show that the ``allowed hopping terms" introduced above which break no symmetry could account for the evolution of the M point alone. Thus a linear combination of the extended $s$-wave bond nematic and the extra hopping could account for the low-temperature Fermi surface and are mathematically equivalent to our unidirectional nematic bond order. Finally we reproduce the unidirectional nematic bond order parameter from the main text. 

\textit{On-site vs bond order: }In a 5-orbital model, one could choose an order parameter $(n_{xz}-n_{yz})\cos{(k_x)}$ which would reproduce the experimental Fermi surface. Since the 5-orbital model does not distinguish between on-site number and inter-site orbital hopping operators, this leaves ambiguity in the interpretation. However when using the full 10-orbital model, the order parameter must be placed in the inter-site hopping terms to reproduce the experimental Fermi surface. The interpretation of on-site stripe antiferro-orbital ordering (or any other pattern) is forbidden since this would require translational symmetry-breaking and additionally has the wrong periodicity. 

\textit{The $d_{xy}$ band: }Experimentally, the sections of Fermi surface with $d_{xy}$ character also shift, although relatively less. We accounted for this with a 10 meV downwards shift of the $d_{xy}$ orbital energy in the tight-binding model. One way to understand this is by noting that the total volume of the Fermi surface must be approximately conserved in the nematic phase since the distortion of the hole pocket is approximately symmetry-preserving. Therefore a downward shift of the $d_{xy}$ electron band dispersion could simply be a necessary consequence of the upward shift of both the $d_{xz}$ and $d_{yz}$ bands. However alternative interpretations could be possible. 

\textit{Dirac points: }Finally we note that our model indicates that spin-orbit coupling gaps out the $d_{xy}-d_{xz}$ band crossing slightly away from M, which has elsewhere been proposed as a Dirac point. This can also be directly observed in the data, where band discontinuities are observed at the crossing points of $d_{xz}$ and $d_{xy}$ bands, e.g. in Fig.~\ref{fig:sm2}. In fact the spin-orbit induced splitting is of comparable magnitude to the Fermi energy and therefore we question whether the Dirac description can be appropriate for these points.

\end{document}